\documentclass[acmsmall]{acmart}

\AtBeginDocument{%
  }

\usepackage{graphicx}
\usepackage{subfigure}
\usepackage{multirow}
\usepackage{amsmath,siunitx,booktabs}
\usepackage{framed}
\usepackage{float}
\usepackage{url}
\usepackage[ruled,vlined,linesnumbered]{algorithm2e}
\usepackage{xspace}
\usepackage{enumitem}
\usepackage{listings}
\usepackage{subcaption}
\usepackage{amsthm}
\usepackage{xcolor}
\usepackage{caption}
\definecolor{shadecolor}{gray}{0.9}

\theoremstyle{definition}
\newtheorem{definition}{Definition}[section]

\newcommand\black[1]{\textcolor{black}{#1}}

\definecolor{darkblack}{RGB}{0,0,139}

\setcopyright{cc}
\setcctype{by}
\acmDOI{10.1145/3808135}
\acmYear{2026}
\acmJournal{PACMSE}
\acmVolume{3}
\acmNumber{FSE}
\acmArticle{FSE128}
\acmMonth{7}
\acmSubmissionID{fse26mainb-p395-p}
\received{2026-02-25}
\received[accepted]{2026-03-24}

\begin{document}

\title{QuanForge: A Mutation Testing Framework for Quantum Neural Networks}

\author{Minqi Shao}
\orcid{0009-0001-4832-5256}
\affiliation{%
  \institution{Kyushu University}
  \city{Fukuoka}
  \country{Japan}
}
\email{shao.minqi.229@s.kyushu-u.ac.jp}

\author{Shangzhou Xia}
\orcid{0009-0006-2775-9633}
\affiliation{%
  \institution{Kyushu University}
  \city{Fukuoka}
  \country{Japan}
}
\email{xia.shangzhou.218@s.kyushu-u.ac.jp}

\author{Jianjun Zhao}
\authornote{Corresponding author.}
\orcid{0000-0001-8083-4352}
\affiliation{%
  \institution{Kyushu University}
  \city{Fukuoka}
  \country{Japan}
}
\email{zhao@ait.kyushu-u.ac.jp}

\setlength{\textfloatsep}{10pt plus 2pt minus 2pt} 

\begin{abstract}
With the growing synergy between deep learning and quantum computing, Quantum Neural Networks (QNNs) have emerged as a promising paradigm by leveraging quantum parallelism and entanglement. However, testing QNNs remains underexplored due to their complex quantum dynamics and limited interpretability. Developing a mutation testing technique for QNNs is promising while requires addressing stochastic factors, including the inherent randomness of mutation operators and quantum measurements. To tackle these challenges, we propose QuanForge, a mutation testing framework specifically designed for QNNs. We first introduce statistical mutation killing to provide a more reliable criterion. QuanForge incorporates nine post-training mutation operators at both gate and parameter levels, capable of simulating various potential errors in quantum circuits. Finally, a mutant generation algorithm is formalized that systematically produces effective mutants, thereby enabling a robust and reliable mutation analysis. Through extensive experiments on benchmark datasets and QNN architectures, we show that QuanForge can effectively distinguish different test suites and localize vulnerable circuit regions, providing insights for data enhancement and structural assessment of QNNs. We also analyze the generation capabilities of different operators and evaluate performance under simulated noisy conditions to assess the practical feasibility of QuanForge for future quantum devices.
\end{abstract}




\maketitle

\section{Introduction}
\label{sec:introduction}

\black{Quantum computing has made rapid progress in recent years, both in terms of hardware capabilities and algorithmic development. One of the most promising areas is quantum machine learning (QML)~\cite{qml}, which combines quantum computing with machine learning to improve mutual performance. In QML, quantum neural networks (QNNs)~\cite{qcnn} have attracted attention for their ability to integrate quantum properties, such as superposition and entanglement, into neural networks. These properties offer potential advantages, including speedup in time complexity~\cite{timecomplexity}. QNNs have achieved initial success in tasks such as image classification~\cite{qnn_survey}, generative models~\cite{quantumgen}, and natural language processing~\cite{quantumnlp}.}

\black{To develop practical applications of QNNs on future quantum devices, concerns about their correctness and reliability, such as adversarial robustness~\cite{qaml}, have become increasingly prominent. These issues highlight the need for systematic testing and verification techniques specifically designed for QNNs. However, such testing frameworks are currently lacking. Designing them is challenging due to the black-box nature of QNN internals and the limited interpretability arising from complex quantum dynamics.}

\black{To make initial efforts, it is natural to consider adapting mature testing techniques from traditional software. Mutation testing (MT)~\cite{mt} is a widely used method for assessing the quality of test data. 
\black{It can provide a systematic way to generate faulty variants and assess whether a test suite can detect injected behavioral deviations.}
MT has been transferred to deep neural networks (DNNs), where various mutation operators~\cite{deepmutation,deepcrime,muff} have been proposed at different levels, including source code, neurons, and layers. 
Meanwhile, several mutation analysis tools~\cite{fortunato2022mutation,muskit,mutationoperator} have also been developed for quantum programs (QPs), focusing on quantum-specific mutation characteristics such as gate-level, circuit-level, and algorithm-level transformations.} 
\black{As an intersection of DNNs and QPs, MT should hold significant promise for improving the behaviors and robustness of QNNs. Compared with the raw accuracy metric, MT can not only quantify the fault-detection capability of a test suite but also reveal fragile regions of the model where small mutations can cause significant changes in output. The MT results could further provide insights for strengthening data quality in a targeted manner and improving model robustness.}

\black{Despite its promise, applying MT to QNNs remains challenging.} Mutation strategies for DNNs are broadly categorized into pre-training and post-training paradigms.
Since QNNs lack a unified paradigm for overall architecture and data encoding, devising a generalizable pre-training strategy is challenging.
Moreover, MuFF~\cite{muff} indicates that the post-training paradigm exhibits superior performance with respect to both resource efficiency and mutant stability. Considering the high experimental cost of QNNs, we prioritize adopting the post-training strategy. However, due to structural differences between DNNs and QNNs, existing post-training mutation methods cannot be directly applied.
\black{On the other hand, mutation techniques for QPs mainly focus on structural changes to circuits, which are insufficient for parameterized gates in QNNs.} 
\black{Manually configuring mutation operators~\cite{muskit} is also inefficient for scalable QNNs.} 
\black{Besides operator design}, existing approaches lack a systematic analysis of mutant quality, i.e., whether mutants are unkillable (cannot be killed by any dataset) or trivial (killed even by the weakest dataset)\black{, leading to low-quality mutants and inaccurate results. Moreover, test oracles \cite{ali2021assessing} based on a single input-output pair can yield aggressive mutant killing for QNN mutants due to their larger-scale datasets.}

Another key challenge is the stochasticity of mutation testing involved in the quantum context, which arises from two aspects: (1) randomness of operator application (e.g., which gates to mutate) and (2) randomness of the measurement process, where repeated executions of a QNN on the same dataset may yield different outcomes. The latter is unique to quantum computing and directly impacts mutation killing and mutation scores. Taking these factors into account, there is a pressing need to develop an efficient and reliable mutation testing framework for QNNs.

To address these challenges, we propose \textbf{QuanForge}, a post-training mutation testing framework to evaluate the quality of test data and the structural robustness of QNNs in a statistical way. \black{First, we define \black{three mutant evaluation metrics} to account for measurement randomness and obtain reliable judgments. Second, by analyzing the functional roles of quantum gates, we design nine mutation operators (MOs) at both gate and parameter levels, covering diverse perspectives and granularity. 
\black{These operators introduce various structural faults or parameter deviations into the circuits, which could, to some extent, simulate several realistic errors and noise during quantum compilation and execution.} 
Finally, QuanForge incorporates stability checking and effectiveness analysis into the mutant generation algorithm to ensure high-quality mutants. It also employs a binary search strategy to automatically adjust operator configurations, maximizing effective mutant generation while avoiding suboptimal manual settings.}
To validate effectiveness, we conducted extensive experiments using benchmark datasets and diverse QNN architectures. QuanForge demonstrates its utility in evaluating the quality of test suites and identifying structural weaknesses by exposing different circuit parts to mutation. \black{Based on observations, several insights are provided for future model enhancement.} We further analyzed the capabilities of different operators to generate mutants \black{and conduct a sensitivity analysis of two hyperparameters}. Additionally, to simulate more realistic quantum execution, we evaluated QuanForge on noisy simulators to provide evidence for its practical value.
Our contributions include:
\begin{itemize}
    \item We introduce statistical mutation killing for QNNs to address the randomness of quantum measurement and provide a more reliable mutant analysis.
    \item We design nine gate-level and parameter-level mutation operators based on quantum gate and circuit characteristics. These post-training operators enable efficient mutation and simulate various errors during circuit design\black{, compilation}, and execution.
    \item We implement a comprehensive mutant generation framework that automatically adjusts operator configurations and integrates a two-step checking procedure to select killable and non-trivial mutants.
    \item We validate QuanForge through extensive experiments on benchmark datasets and QNNs. The results demonstrate its effectiveness in evaluating the quality of the test suite and analyzing the model robustness by mutating different target regions and gate types.
\end{itemize}

\section{Background and related work}
\label{sec:background}

\subsection{Quantum Computing}
\label{sec:quantumcomputing}

\noindent \textbf{Qubits.} A quantum bit, or qubit, is the fundamental unit of information in quantum computing. Unlike classical bits, qubits can exist in a \textit{superposition} of computational basis states. A pure state is written as $|\phi\rangle = \alpha|0\rangle + \beta|1\rangle$ with $\alpha,\beta \in \mathbb{C}$ as probability amplitudes. Upon measurement, the qubit collapses to $|0\rangle$ with probability $|\alpha|^2$ and to $|1\rangle$ with probability $|\beta|^2$.

\noindent \textbf{Quantum gates and circuits.} 
Quantum gates are essential components in quantum programs that perform rotation or change entanglement on qubits. 
Quantum circuits consist of sequences of qubits and gates, 
and realize functionalities by modifying the selection and parameterization of quantum gates, the choice of target qubits, and the order of gate execution.



\noindent  \textbf{Quantum measurement.} Quantum measurement projects a superposition into a definite classical state according to probability amplitudes, with the state collapsing irreversibly. This operation is conventionally performed at the terminal stage of a quantum circuit with specific basis. 
Due to the probabilistic nature, 
\black{reliable evaluation of a quantum circuit often requires multiple measurements.}


\subsection{Quantum Neural Networks}
\label{sec:qnn}

QNNs, inspired by DNNs, are typically built from parameterized quantum circuits (PQCs) with predefined structures and tunable gate parameters. 
A typical QNN consists of three components: a data encoding layer that maps classical data to quantum states, a parameterized circuit layer that extracts features via quantum state transformations, and a measurement layer that extracts classical information from closed quantum systems.

Various QNN variants have been proposed~\cite{hqnn, qcnn, qcl, hcqc} recently. Based on circuit functionality, current QNNs can be categorized into three types: \textit{circuit-body QNNs} \cite{qcl,qcnn}, which use medium-sized PQCs as the backbone; \textit{circuit-kernel QNNs} \cite{kernelqnn}, which employ PQCs as convolutional kernels sliding for feature extraction; and \textit{hybrid QNNs} \cite{hqnn}, which integrate PQCs with classical layers, where PQCs serve as either a preprocessing or output layer. 

For the data encoding layer, two approaches are commonly used. \textit{Amplitude encoding} encodes data features as amplitudes of a quantum state, which require relatively few qubits but deep circuits to implement. \textit{Angle encoding} encodes features as rotation-gate parameters, enabling efficient implementation but consuming more qubits. For the parameterized circuit layer, two representative designs are \textit{block stacking} \cite{qcl}, which repeatedly stacks the same block, and \textit{hierarchical structures} \cite{qcnn}, which reduce circuit freedom by measuring subsets of qubits as circuit deepens. \black{Finally, given a finite number of shots, measurement collapses the quantum state, and expectation values on selected qubits are used as QNN outputs~\cite{hcqc}.}



\subsection{Mutation Testing for Classical Neural Networks}
\label{sec:mt4dnn}

In recent years, various mutation testing techniques~\cite{deepmutation,deepcrime,muff} have been proposed for DNNs, which assess test adequacy by injecting artificial faults to create mutants and evaluate test effectiveness.

One category is the post-training mutation~\cite{munn,deepmutation,muff}, which directly modifies the weights or neurons of the trained models. DeepMutation~\cite{deepmutation} defines model-level operators that target DNN structures such as neurons and layers. It is efficient but may introduce large changes, leading to significant performance degradation. Moreover, weight- or neuron-level operators have limited interpretability and cannot mimic real faults in DL systems. With manually defined operator parameters, the variability and instability of mutants have posed a threat. To address these, MuFF~\cite{muff} introduces a stability check mechanism and two novel operators to generate stable and sensitive mutants. 
Another category is the pre-training mutation~\cite{deepcrime}, where faults are injected into source programs before training. It targets various components 
\black{involved during the training pipeline, including training data, hyperparameters, activation function, and loss function}.

Beyond mutation operator design, Jahangirova and Tonella~\cite{20emp} proposed statistical methods for mutation killing, taking into account the nondeterminism of model training. To cover more configurations of operator parameters, some works~\cite{20emp,deepcrime} further treated each configuration as a potential killing target and computed the killed configurations with respect to the training data.




\black{While both are used to assess test adequacy, MT for QNNs differs from that for DNNs in:}

\black{
\noindent $\bullet$ \textit{Operator scope.} In DNNs, post-training operators defined for model internals are not directly transferable to QNNs, since components such as neurons, layers, and activation functions do not have clear counterparts in PQC-based QNNs. Meanwhile, source-level or pre-training mutations are often impractical, as each mutant requires retraining from scratch, which is costly and can be unstable under stochastic quantum execution.
}

\black{
\noindent $\bullet$ \textit{Evaluation.} QNNs involve additional stochastic factors due to probabilistic measurement and hardware-related noise (e.g., coherent errors), so mutation killing cannot rely solely on a single prediction. Instead, it needs to consider more general model behavior (e.g., accuracy) under repeated sampling. Correspondingly, mutant-quality evaluation (e.g., filtering unkillable or trivial mutants) needs to be redefined for QNNs to prevent budget waste and improve testing efficiency.
}

\subsection{Mutation Analysis and Testing for Quantum Programs}
\label{sec:mt4qp}

Testing quantum programs has attracted increasing attention recently. Previous work on quantum program testing has examined the basic challenges of this field~\cite{miranskyy2019testing}, proposed systematic testing frameworks~\cite{long2024testing}, and developed various testing techniques~\cite{honarvar2020property,quratest,quantumconcolic,long2024equivalence,long2025black,li2025preparation,wang2021generating,jin2025novaq,li2020projection,wang2021application}.

Mutation analysis has been adopted to assess the quality of test suites for quantum programs. Muskit~\cite{muskit} defines gate-level mutation operators, focusing on the number and location of the gate \black{manually defined}. QMutPy~\cite{fortunato2022mutation} targets Qiskit programs and defines operators for measurement and gate construction, using syntactically equivalent gates.
A large-scale empirical study~\cite{mutationoperator} compares the detection of faults and the cost effectiveness of various mutation strategies across circuit types and algorithms. QCRMut~\cite{qcrmut} proposes four types of operators that preserve circuit structure, using gate equivalence and randomized mutation to generate meaningful but minimally disruptive changes.
For mutation killing, two test oracles~\cite{ali2021assessing} are widely used in QP testing: \textit{WOO} (Wrong Output Oracle), which checks whether the native output changes, and \textit{OPO} (Output Probability Oracle), which considers probabilistic behaviors by repeated execution.





\black{
Compared to these QP-oriented approaches, MT for QNNs differs in:
}

\black{
\noindent $\bullet$ \textit{Purpose.} Both lines of work are commonly used to assess test adequacy. MT for QPs generates faulty variants by injecting code-level bugs in program implementations, and is often used to evaluate the fault-detection capability of testing techniques.
MT for QNNs injects perturbations to emulate model-level weaknesses arising from training data quality, circuit architecture, and optimization dynamics, and can also help identify sensitive regions for model enhancement \cite{mt4faultlocate}.}

\black{
\noindent $\bullet$ \textit{Mutant generation.} Gate-level mutation for QPs typically supports insertion, deletion, and replacement, and some approaches also mutate measurements. The latter is unsuitable for QNNs because QNN outputs are directly derived from measurements, whose mutation destroys the output definition rather than injecting internal faults.
Additionally, finer-grained parameter-level mutations can lead to significant behavioral changes in QNNs due to their parameter-learning mechanism. 
Therefore, MT for QNNs should include a more diverse gate-level and parameter-level operators with automatically adjusted configurations.}

\black{
\noindent $\bullet$ \textit{Evaluation.} QP mutation killing is often decided by per-input oracle judgments \cite{ali2021assessing}. For dataset-driven QNNs, relying only on a single input-output pair may overestimate the killing capability of datasets. Mutation killing should be defined based on general model behavior, such as accuracy, while accounting for measurement randomness in QNN outputs.
Moreover, beyond mutation score, richer metrics are needed to characterize structural sensitivity (e.g., to gate type, circuit depth, or circuit region) and to provide actionable insights for model improvement.
}


\section{Method}
\label{sec:method}

\begin{figure}[t]
    \centering
    \includegraphics[width=0.85\linewidth]{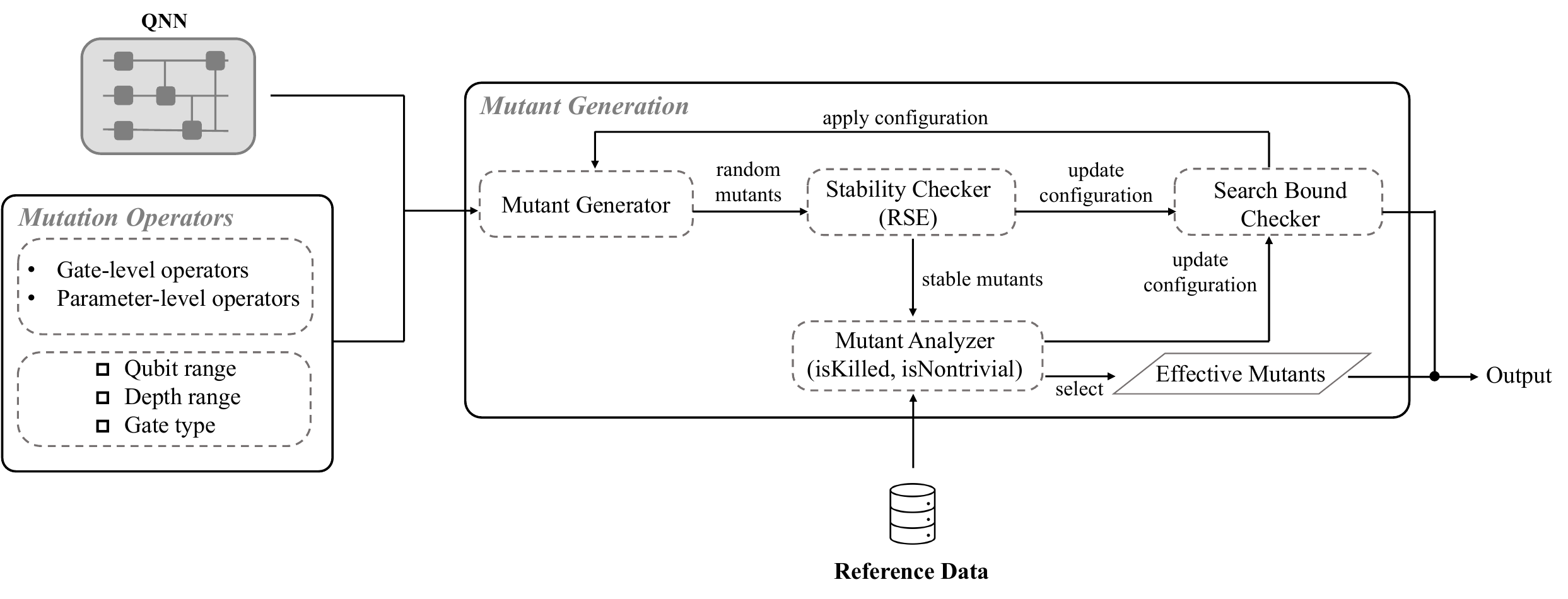}
    \vspace{-3mm}
    \caption{The overall workflow of effective mutant generation in QuanForge.}
    \label{fig:workflow}
    \vspace{-3mm}
\end{figure}

\black{Figure \ref{fig:workflow} illustrates the overall workflow of QuanForge, which consists of two major modules: (1) the \textit{Mutation Operators} module (Section \ref{sec:operators}) contains nine gate-level and parameter-level mutation operators, together with user-specified configurations for the following process, and (2) the \textit{Mutant Generation} module (Section \ref{sec:mutant_generation}) contains the process of mutant initialization, checking, and analysis. Specifically, the \textit{mutant generator} first takes the original QNN and the defined operators as input and initially produces a batch of random mutants. Based on the Relative Standard
Error (RSE) criterion, stable mutants are obtained by a \textit{stability checker}. Subsequently, the \textit{mutant analyzer} evaluates these mutants in terms of killability and nontriviality and selects effective mutants.}
During the checking and analyzing procedures, the specific configuration of an operator parameter is updated based on the evaluation results of mutants at the same time. The updated configuration is then passed to a \textit{search-bound checker}, which decides whether the generation process should proceed or terminate by returning all effective mutants so far.

\subsection{\black{Mutant Evaluation Metrics}}
\label{sec:definitions}

\black{As illustrated in \cite{deepcrime,muff}, the training process of DNNs is inherently nondeterministic, and this still holds for QNNs due to their similar training procedures. However, unlike DNNs whose outputs are deterministic during inference, QNNs exhibit an additional aspect of randomness arising from quantum measurement. 
In each measurement, a qubit collapses probabilistically to a definite basis state, and the average over multiple measurements corresponds to the expectation value, which constitutes the QNN outputs. With few measurements, the value is more likely to deviate from the theoretical value computed by matrix simulation. 
To ensure the output reliability, we calculate the number of measurements
required to achieve a specific confidence level and an error bound based on Hoeffding's inequality \cite{inequality}. A QNN output obtained with $s$ measurements is considered as one \textit{prediction}. 
However, repeated predictions on the same input can still exhibit varying performance in subsequent task-specific processing, making it unreliable to determine whether a mutant should be killed based on a single input. 
Hence, the test oracle for QPs that focuses on a single input-output pair may lead to overestimation when applied to large-scale datasets. To address this, we incorporate \textit{reference data}, which are independent of the test data,
into the evaluation of mutant performance. Ideally, such data should achieve high accuracy with the original model while remaining sensitive to the decision boundary, enabling it to distinguish the original model from its mutants. Considering this, we adopt training data as reference data. In this section, we propose three metrics for evaluating mutants with respect to their stability, killability, and non-triviality.}


\black{Due to the inherent randomness of the mutation operator (e.g., the random selection of target gates), even mutants generated from the same operator and configuration can exhibit unstable performance. As a result, among a fixed number of mutants, many might experience significant accuracy degradation, undermining generation efficiency.} To address this instability, we adapt the Relative Standard Error (RSE)~\cite{muff} to quantify the variance in performance across a batch of mutant instances.
Let $M_s$ denote a set of random mutants generated with a given mutation operator configuration. By evaluating each mutant $M$ in $M_s$ using reference data, we obtain accuracies $A_{M_s}=\langle A_1^\prime, ..., A_{|M_s|}^\prime \rangle$ with mean $\mu$ and standard deviation $\delta$. \textcolor{black}{RSE score} is defined as: 

\vspace{-1mm}
{\small
\begin{equation}
\label{eq:rse}
    RSE = \frac{\delta}{\mu\sqrt{|M_s|}}
\end{equation}
\vspace{-1mm}
}

\black{RSE measures the variability of an estimate relative to its mean, where a smaller value indicates better stability.}
\black{As $\mu$ and $\delta$ are bounded, the RSE score converges toward zero with the increasing number of mutants, thereby ensuring the eventual acquisition of a stable mutant set.}
To achieve a smaller RSE score, we can explore the generation space of a configuration more adequately by collecting more mutants compared to fixed settings.

\begin{definition}[\textit{Stable mutants}]
    \black{A set of mutants is considered stable if, on reference data, their RSE score is smaller than a threshold $\tau_{RSE}$.}
\end{definition}

\black{The randomness of MO application may not only generate mutants with degraded performance, but also lead to mutants that show minimal differences from the original model.} Such mutants can lead to inefficiencies in the testing process \black{since they cannot be killed by any dataset}. Therefore, we employ the killable mutant metric to filter out mutants that exhibit only minor variations.

Given the inherent unreliability of a single prediction arising from quantum measurement, we perform the evaluation based on the performance distributions obtained from repeated predictions.
We adopt a generalized linear model (GLM) to evaluate whether a significantly different distribution is introduced by the mutant based on the $p$-value, and use Cohen's $d$ to quantify the effect size. 
\black{Specifically, the \textit{dependent variable} is the model accuracy and the \textit{independent variable} is whether the model is original or mutated. We fit a GLM with a Gaussian exponential family and identity link. The null hypothesis is that the regression coefficient is zero, i.e., the original and mutated models have similar performance.} 
Formally, given the original model $O$, a mutant $M$, reference dataset $D$, and two thresholds ($\alpha$,$\beta$), the prediction is repeated in $O$ and $M$ for $n$ times, respectively, generating two accuracy distributions, $A_{O}=\langle A_1, ..., A_n \rangle$ and $A_{M}=\langle A_1^\prime, ..., A_n^\prime \rangle$. Then, the statistical mutation killing $isKilled$ is defined as follows, which means that a mutant can be killed only if its accuracy distribution is significantly different from the original:

\vspace{-1mm}
{\small
\begin{equation}
\label{eq:iskilled}
    isKilled(O,M,D,n) = 
    \begin{cases}
        \textit{True}, &
           \begin{aligned}
               &\text{if } p\_value(A_O(D),A_M(D)) < \alpha \\
               &\text{ and } effectSize(A_O(D),A_M(D)) \geq \beta
           \end{aligned} \\
        \textit{False}, & \text{otherwise}
    \end{cases}
\end{equation}
\vspace{-2mm}
}

\begin{definition}[\textit{Killable mutant}]
    A mutant is considered killable if it can be killed by reference data according to $isKilled$.
\end{definition}

\black{An opposite issue to killable mutants is that some mutants may exhibit vast differences from the original model. Such mutants can be detected by almost all data, which makes them ineffective in assessing the quality of different test data. We use the non-trivial mutant metric to filter out such mutants with overly large deviations, depending on their average accuracy over multiple predictions. Formally, given a mutant $M$, reference dataset $D$, and a threshold $\tau_{trivial}$, after obtaining the accuracy distribution $A_{M}=\langle A_1^\prime, ..., A_n^\prime \rangle$, the $isNontrivial$ is defined as:}

\vspace{-1mm}
{\small
\begin{equation}
    \label{eq:isNontrivial}
    isNontrivial(M,D,n) = 
    \begin{cases}
        \textit{True}, & \text{if } \frac{1}{n}\sum_{i=1}^n A_i ' \geq \tau_{trivial}\\
        \textit{False}, & \text{otherwise}
    \end{cases}
\end{equation}
}
\vspace{-2mm}

\begin{definition}[\textit{Non-trivial mutant}]
A mutant is considered non-trivial if its average accuracy is larger than a threshold $\tau_{trivial}$. 
\end{definition}

\black{In summary, mutant stability characterizes the general performance of a batch of mutants and can be further regarded as a property of the mutation operator. It reflects the capability of this operator to generate stably behaving mutants with a specific configuration. The killability and non-triviality are properties evaluated for each individual mutant instance. We define mutants that simultaneously satisfy both killable and non-trivial properties as \textit{effective mutants}. Such mutants exhibit sufficient differences to be detected by test data while also distinguishing the quality of different test suites.}

\subsection{Post-training Mutation Operators for QNNs}
\label{sec:operators}

Mutation operators (MOs) have been widely studied for DNNs, but cannot be directly applied to QNNs due to the intrinsic properties of quantum computing. Unlike DNNs composed of neurons connected by weights, QNNs are constructed from qubits and quantum gates, endowed with the properties of superposition and entanglement. The novel structure of quantum circuits renders the neuron- and weight-based mutation operators in DNNs inapplicable. Moreover, the inefficiency and resource overhead of retraining QNNs also prevent a direct adaptation of source-level mutation strategies from DNNs. To accommodate the unique structure of QNNs, we extend the QP-oriented MT techniques and propose two granularities of MOs at the gate and parameter levels. These operators enable a more fine-grained mutation of quantum gates with respect to positional information, parameter values, and control dependencies.

Furthermore, we leverage some configurations to impose constraints when applying MOs. 
\textit{Qubit range} (which qubits can be mutated) and \textit{depth range} 
\black{(the gate sequence length along the qubits)}
are utilized to describe a target mutation scope in QNNs, ensuring that mutation occurs solely within this scope. \black{\textit{Gate type} restricts mutations to specific types of gates as categorized in Table \ref{tab:gate_category}, and \textit{gate percentage} determines the total number of gates to mutate.}

\begin{table}[t]
\centering
\caption{Common quantum gates used in QNNs}
\label{tab:gate_category}
\resizebox{0.7\linewidth}{!}{
\tiny
\begin{tabular}{@{}cll@{}}
\toprule
Criteria & \multicolumn{1}{c}{Type} & \multicolumn{1}{c}{Name} \\ \midrule
\multirow{6}{*}{Gate Function} & Controlled & CX, CZ, CNOT, CRX, CRY, CRZ, ControlledPhaseShift \\
 & Hadamard & Hadamard \\
 & Pauli & PauliX, PauliY, PauliZ, CX, CZ \\
 & Phase & PhaseShift, ControlledPhaseShift \\
 & Rotation & RX, RY, RZ, Rot, U3, CRX, CRZ, CRY \\
 & Swap & SWAP, CSWAP \\ \midrule
\multirow{3}{*}{Gate Size} & Single & Hadamard, PauliX, PauliY, PauliZ, PhaseShift, RX, RY, RZ, Rot, U3 \\
 & Two & CX, CZ, CNOT, SWAP, CRX, CRY, CRZ, ControlledPhaseShift \\
 & Multi & CSWAP \\ \bottomrule
\end{tabular}}
\end{table}

\subsubsection{Gate-level mutation operators}
\label{sec:gateoperator}
Quantum gates, as the fundamental building blocks of quantum circuits, enable the construction of QNNs with substantially different functionalities by applying them to different qubits at various positions. 
\black{Thus, during inference, the misuse of quantum gates can induce noticeable performance degradation, typically manifested as an accuracy drop. Beyond syntactic programming mistakes, gate-level mutations can serve as a proxy for quantum-unique faults in realistic execution, which can disrupt the model behavior. For example, SWAP insertion \cite{mapping} may be introduced by hardware connectivity constraints and compilation, and additional gates could be brought by suboptimal decompositions \cite{gatedecompose}.} 
To simulate these issues, we propose the following six gate-level mutation operators:


\begin{figure}[t]
    \centering
    \includegraphics[width=0.85\linewidth]{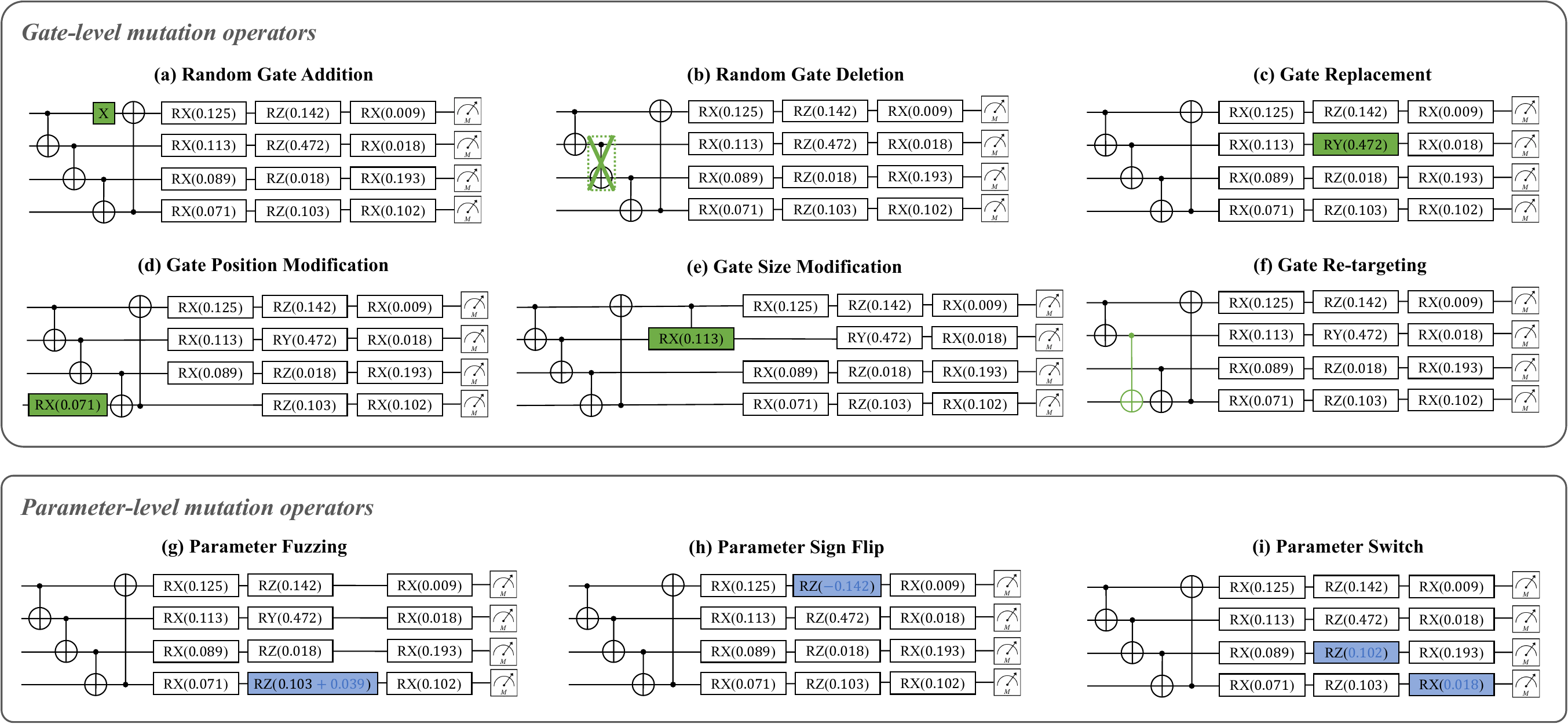}
    \vspace{-2mm}
    \caption{Examples of mutants generated by different mutation operators.}
    \label{fig:mutant_example}
\end{figure}

\textbf{Random Gate Addition (RGA):} This operator adds a fixed number of gates, selected from the specified types, at random positions within the defined mutation scope. For parameterized gates, parameters are randomly generated from a predefined distribution, such as a Gaussian or exponential distribution.
Figure~\ref{fig:mutant_example}(a) shows an X gate added to qubit 0 at position 2.

\textbf{Random Gate Deletion (RGD):} 
This operator removes a fixed number of gates at random positions within the defined mutation scope. 
Figure~\ref{fig:mutant_example}(b) shows an example in which a CNOT gate is deleted on qubits (1, 2) at position (2, 1).

\textbf{Gate Replacement (GR):} 
Combining the notions of gate equivalence~\cite{qmutpy,qcrmut}, this operator replaces a gate with another of the same size and parameter count to preserve the circuit structure and the position of the gate. GR ensures the syntactic validity and circuit structural similarity of the generated mutant, while introducing variations in gate functions.
Figure~\ref{fig:mutant_example}(c) shows the RX gate on qubit 1 at position 4 replaced with an RY gate, keeping the rotation parameter unchanged.

\textbf{Gate Position Modification (GPM):} Since the arrangement of quantum gates can also affect the execution results, this operator shifts a gate to a different location along its assigned qubit, while keeping the applied qubit unchanged.
From a model perspective, GPM disrupts the non-commutative structure of QNNs and changes how parameters affect the output. Figure~\ref{fig:mutant_example}(d) shows an RX gate on qubit 3 moved from position 3 to position 1.

\textbf{Gate Size Modification (GSM):} 
This operation modifies the size of the original gate by adding or removing the control relations of the qubit. GSM directly affects the structure of QNN entanglement, thereby affecting the entangled features and the overall trainingability of the model~\cite{qleet}.
Figure~\ref{fig:mutant_example}(e) shows a new CRX gate on qubit 1 at position 3 with an additional control relation of qubit 0.

\textbf{Gate Re-targeting (GRT):} This operator modifies the target qubit(s) of a quantum gate. 
Since each qubit encodes distinct features and participates in different entanglement patterns, GRT can disrupt inter-qubit interactions and shift feature extraction behaviors. 
Figure~\ref{fig:mutant_example}(f) shows a CNOT gate on qubits (1, 2) at position (2, 1) re-targeted to qubits (1, 3), with gate type preserved.

\subsubsection{Parameter-level mutation operators}
\label{sec:paramoperator}
The parameters of parameterized gates are the only trainable components in a QNN and play a crucial role in feature extraction and the formation of decision boundaries, much like weights in DNNs. Parameter-level mutation simulates optimization issues such as underfitting or overfitting due to suboptimal parameters, which are especially relevant in QNNs that suffer from barren plateaus \cite{barren}. Moreover, it can provide an abstraction of realistic hardware noise, where limited control precision can induce inaccurate gate operations~\cite{coherenterror}.
To retain the circuit structure while exerting a non-negligible influence on the model behavior, we propose the following three finer-grained parameter-level mutation operators:

\textbf{Parameter Fuzzing (PF):} This operator perturbs a gate by injecting random noise into its parameter, sampled from predefined distributions such as Gaussian, uniform, or exponential. The injected noise induces deviations in gate operations, impairing feature extraction.
The effectiveness of mutants depends on the noise scale. With mild noise, the killability of mutants cannot be ensured, whereas strong noise may result in trivial mutants.
Figure~\ref{fig:mutant_example}(g) shows an RZ gate on qubit 3 at position 4 with its parameter perturbed by 0.039.

\textbf{Parameter Sign Flip (PSF):} \black{This operator flips the parameter sign of the target gate, thereby reversing its functional effect on the qubit.} 
\textcolor{black}{From the perspective of quantum computing, PSF induces a rotation of the quantum state in the opposite direction, which thus differs from the perturbation introduced by PF.}
In Figure \ref{fig:mutant_example}(h), the parameter of the RZ gate is flipped to -0.142.

\textbf{Parameter Switch (PS):} \black{This operator switches the parameters of two gates belonging to the same category, analogous to Neuron Switch~\cite{deepmutation} in DNNs. PS perturbs the original non-commutative structure of the QNN, thereby affecting feature transformation.} 
\textcolor{black}{Figure \ref{fig:mutant_example}(i) shows the parameter switch between an RZ gate on qubits 2 and an RX gate on 3.}


\subsection{\black{Mutant generation}}
\label{sec:mutant_generation}

\begin{algorithm}[t]
\footnotesize
\DontPrintSemicolon
\caption{Generate Effective Mutants}
\label{alg:generate}
\KwIn{Original model $O$, reference dataset $D$, mutation operator $MO$, max iteration $k_{\max}$, number of mutants generated each iteration $i$, search range for gate percentage $[lb,ub]$}
\KwOut{Effective mutants $results$}
\SetAlgoSkip{smallskip} 
\SetArgSty{textnormal}   
\SetFuncSty{textnormal}
\SetKwFunction{FBS}{GenerateEffectMutants}
\SetKwProg{Fn}{def}{:}{}
\SetKwComment{tcp}{\textcolor{darkblack}{//~}}{}
$results \gets \emptyset$\;
\Fn{\FBS{$lb, ub,i$}}{
    $mid \gets (lb + ub)/2$; \ \ \ \  $k \gets 0$; \ \ \ \  $mut\_list \gets \emptyset$\;
    \While{$k \leq k_{\max}$ and $RSE(mut\_list, D) > \tau_{RSE}$ \hfill \tcp{\textcolor{darkblack}{Stability Checker}}}{
        $k \gets k + 1$\;
        $mutants \gets \textit{GenerateMutant}(O, MO, i, mid)$\;
        $mut\_list \gets mut\_list \cup mutants$\;
    }
    \If{$RSE > \tau_{RSE}$}{
        $ub \gets mid$; \ \ \ \ $i\gets 1.5\times i$\;
    }
    \Else{
        $mut\_acc\_list \gets CalculateAcc(mut\_list,D)$\;
        $ori\_acc\_list \gets Copy(CalculateAcc(O,D))$\;
        \If{$isSetKilled(ori\_acc\_list,mut\_acc\_list)$ \hfill \tcp{\textcolor{darkblack}{whether the config generates killable mutants}}}{ 
            $ub \gets mid$\;
            $results \gets results \cup  SelectEffectiveMutants(O,D,mut\_list)$ \hfill \tcp{\textcolor{darkblack}{Mutant Analyzer}}
            }
        \Else{
            $lb \gets mid$\;
        }
    }
    \If{$[lb,ub]$ cannot be refined \hfill \tcp{\textcolor{darkblack}{Search Bound Checker}}}{
        \KwRet $results$\;
    }
    \Else{
        \KwRet \FBS{$lb, ub, i$}\;
    }
}
\end{algorithm}

Based on the concepts of mutant evaluation metrics in Section~\ref{sec:definitions} and MOs in Section~\ref{sec:operators}, in this section, we introduce the \textit{Mutant Generation} module, which aims to produce effective mutants for future evaluation and analysis.

\textbf{Mutant Generator:}
\black{Given the target QNN model and a preselected MO strategy, this module generates a substantial set of random mutants within the mutation scope defined by related configurations. The generated mutants are subjected to further property analysis, from which the subset of effective mutants will be identified.}

\textbf{Stability Checker:}
\black{This module aims to generate a stable collection of mutants using RSE. Since RSE decreases with the number of mutants as in Equation \ref{eq:rse}, we adopt an iterative generation process during which new random mutants are gradually produced until RSE falls below $\tau_{RSE}$ or a maximum iteration limit is reached. 
Once RSE reaches the threshold, the resulting set of mutants is considered stable and used for subsequent analysis and selection.}

\textbf{Mutant Analyzer:}
\black{The mutant analyzer selects effective mutants from stable mutants. For each mutant, we perform multiple predictions on the reference data and obtain its accuracy distribution. With the accuracy distribution of the original model as a reference, the killability and nontriviality of the mutant are evaluated based on Equations \ref{eq:iskilled} and \ref{eq:isNontrivial}, respectively.
The mutant will be retained only if the two conditions are both satisfied.}

\textbf{Search Bound Checker:}
\black{The gate percentage directly affects mutant behaviors as a mutation-strength configuration.}
\black{A large percentage injects more errors with stronger randomness, resulting in more unstable and trivial mutants, while a small one keeps mutants close to the original model, but increases the risk of unkillable mutants. To balance this trade-off, we employ binary search to automatically adjust the search bounds. The search goal is to approach a configuration that can generate killable mutants while remaining close to a boundary that renders most mutants unkillable. Updates are guided by feedback from the Stability Checker and Mutant Analyzer. \black{In terms of stability, if a stable RSE cannot be achieved within the maximum iterations, the current configuration is aggressive, and the upper bound is reduced to weaken the mutation impact. For the analyzer, the bounds are reduced if current mutants are killable, promoting a smaller configuration for harder-to-kill mutants; otherwise, they are set to larger values to find a more likely killable configuration.} Finally, after updating, the checker determines whether new bounds allow for further generation. If not, the algorithm terminates and returns all effective mutants generated so far.}

Algorithm \ref{alg:generate} presents the recursive mutant generation process. The inputs include the original model $O$, reference data $D$, and a mutation operator $MO$. The search space for the gate percentage is restricted by a lower bound $lb$ and an upper bound $ub$. $k_{max}$ specifies the maximum iterations for generating stable mutants, and $i$ denotes the number of mutants generated in each iteration. Initially, the configuration of the gate percentage is set as the middle point of the search bounds (line 3). In each iteration, $i$ new mutants are appended to the mutant set by applying $MO$ with the current configuration (lines 7-8). The loop terminates until the RSE is less than $\tau_{RSE}$ or $k_{max}$ is reached (line 4).
An RSE above the threshold reflects the instability of the current configuration, and the search range is narrowed to explore a smaller-value one (line 10). Also, due to the current configuration yielding no stable mutants, the number of mutants generated $i$ is increased to 1.5 times (line 10) for the next generation iteration to ensure a sufficient number of mutants in the end. If RSE is stable enough, the algorithm proceeds to the analysis phase. 
Before analyzing individual mutants, to further improve efficiency, it checks whether the current configuration can generate killable mutants by slightly adapting \textit{isKilled} metric (line 14). Each element in the mutant accuracy list $mut\_acc\_list$ represents the average accuracy over multiple predictions of a mutant (line 12). The original accuracy is replicated into multiple copies to match the length of the mutant accuracy list (line 13). The $ori\_acc\_list$ and $mut\_acc\_list$ are assigned as $A_O$ and $A_M$, respectively.
If true, the upper bound is updated to the middle point, and effective mutants are selected (lines 16-17). Otherwise, the lower bound is set to a larger value (line 19). The search process terminates when the search bounds can no longer be refined (line 20). Finally, we can obtain a set of effective mutants that cover different configurations of gate percentage.

\section{Evaluation}
\label{sec:exp}

\black{We implement QuanForge using PennyLane 0.42~\cite{pennylane} and Pytorch 2.8~\cite{torch}. } \black{All experiments are conducted on systems equipped with Intel Xeon E5-1650 (6 cores, 32GB) and Ubuntu 22.04.} To evaluate the effectiveness of QuanForge, we aim to address the following research questions:

\noindent $\bullet$ \textbf{RQ1}: How effective is QuanForge in assessing the quality of a test suite?

\noindent $\bullet$ \textbf{RQ2}: How sensitive are different parts of QNNs to mutation operators?

\noindent $\bullet$ \textbf{RQ3}: What is the impact of mutating different types of quantum gates on the model performance?


\subsection{Target Datasets and Models}

\begin{table}[t]
\centering
\caption{Dataset and QNN architectures}
\label{tab:dataset}
\resizebox{0.95\textwidth}{!}{
\tiny
\begin{tabular}{llllllll}
\hline
Dataset & Task & Target classes & QNN & Gates & Output qubits & \#gate & Acc (\%) \\ \hline
\multirow{6}{*}{MNIST} & \multirow{4}{*}{Binary classification} & \multirow{4}{*}{digits 0 and 1} & QCL & RX, RZ, CNOT & 0, 1 & 150 & 100 \\
 &  &  & QCNN & PauliX, RX, RY, RZ, U3, CRX, CRZ, CNOT & 0, 2 & 134 & 100 \\
 &  &  & HCQC & RY, CNOT & 7 & 64 & 100 \\
 &  &  & DRNN & RX, RZ, CRZ & 0, 1 & 120 & 99.29 \\ \cline{2-8} 
 & \multirow{2}{*}{Ternary classification} & digits 0, 1 and 2 & QCL & - & 0, 1, 2 & 150 & 91.86 \\
 &  & digits 4, 5 and 7 & QCNN & - & 0, 2, 4 & 134 & 90.16 \\ \hline
\multirow{6}{*}{FashionMNIST} & \multirow{4}{*}{Binary classification} & \multirow{4}{*}{T-shirt and Trouser} & QCL & - & 0, 1 & 150 & 92.50 \\
 &  &  & QCNN & - & 0, 2 & 134 & 93.00 \\
 &  &  & HCQC & - & 7 & 64 & 94.25 \\
 &  &  & DRNN & - & 0, 1 & 120 & 92.21 \\ \cline{2-8} 
 & \multirow{2}{*}{Ternary classification} & T-shirt, Trouser and Pullover & QCL & - & 0, 1, 2 & 150 & 89.67 \\
 &  & Trouser, Pullover and Dress & QCNN & - & 0, 2, 4 & 134 & 91.50 \\ \hline
\end{tabular}}
\vspace{-1mm}
\end{table}

\black{We choose two datasets commonly used in prior QML works for image classification: MNIST and FashionMNIST.} The MNIST dataset~\cite{mnist} contains 70,000 grayscale images of handwritten digits (0–9), each with a resolution of $28 \times 28$. The FashionMNIST dataset~\cite{fashion} has the same format but comprises images from ten categories of clothing items, such as T-shirts and trousers.
 
To evaluate a more diverse set of QNNs, we select four representative architectures with different block-stacking and hierarchical-structure strategies.

\textbf{Quantum Circuit Learning (QCL)}~\cite{qcl} is a classical–quantum hybrid model designed to approximate nonlinear functions through parameterized quantum circuits. \black{QCL belongs to a block-stacking structure and employs amplitude encoding.}

\textbf{Quantum Convolutional Neural Network (QCNN)}~\cite{qcnn} is inspired by classical CNNs, implementing convolution, pooling, and fully connected operations using quantum circuits. It is effective in mitigating the barren plateau problem \cite{qcnnadvantage}. QCNN is hierarchical and employs amplitude encoding.

\textbf{Hierarchical Circuit Quantum Classifier (HCQC)}~\cite{hcqc} is another hierarchical model that reduces the circuit's degrees of freedom as depth increases, using translationally stacked, invariant ansatz blocks. Here we use the ansatz U\_SO4 with amplitude encoding. 

\black{\textbf{Data Re-uploading Neural Network (DRNN)}~\cite{drnn} re-uploads data features as rotation angles to multiple qubits, combined with extra trainable parts to form a universal classifier. It addresses the limited expressivity of a single qubit with fewer quantum computational resources. DRNN adopts a block-stacking structure and angle encoding.}

\black{Considering the time overhead, we adjust the image size of the DRNN to $8\times8$ and configure it with 6 qubits, while the other QNNs are set to 8 qubits} \black{and images are downsampled to $16\times16$.}

To obtain reliable QNN outputs, we need to estimate the minimum number of measurements $s$ to 
\black{achieve an error bound $\epsilon$ with a confidence level of $\delta$. Let $P_i=\{X_i^{(1)},\dots,X_i^{(s)}\}$ be the outcomes of $s$ measurements on $i$-th qubit where $X_i^{(j)}$ is bounded and independent of each other. The average estimate is $\hat{\mu}=\frac{1}{s}\sum_{j=1}^{s}X_i^{(j)}$. The precise QNN outputs are defined as the expectation value of measurement results, denoted as $\mu$. Based on Hoeffding's inequality, for each qubit, we can obtain $Pr(|\hat{\mu}-\mu|\geq \epsilon)\leq 2\text{exp}(-2s\epsilon^2)$. Then the union bound is applied to all qubits and the total probability is bounded by $1-\delta$, producing $2q\text{exp}(-2s\epsilon^2)\leq 1-\delta$. Finally, we obtain $s\geq\frac{1}{2\epsilon^2}\text{log}(\frac{2q}{1-\delta})$, indicating that measurement times scale logarithmically with the number of output qubits, and a similar cost for QNNs with the same output qubits.} 
Given that $q$ is 2 and 3 for binary and ternary classification, respectively, we set $s=1000$ for a prediction to achieve a $\epsilon$ of 0.05 and a $\delta$ of 0.95.

\subsection{Metrics}
\label{sec:metric}

\black{To provide a quantitative measure of the dataset's ability to detect injected faults, we adopt \textit{mutation score} based on statistical killing. To enable a more fine-grained evaluation of test suites, the killing criterion is further refined at the class level~\cite{deepmutation}. Better-quality test suites are expected to achieve higher mutation scores as they are more sensitive to the mutated decision boundary of mutants. The mutation score is calculated as follows:}
\vspace{-1mm}
{\small
\begin{equation}
    MutationScore=\frac{\sum_{m_i\in M_e}|killedClasses(T, m_i)|}{|M_e|\times c}
\end{equation}
\vspace{-1mm}
}

\black{where $T=\{T_1, ..., T_c\}$ denotes the $c$ subsets of test data corresponding to the 
$c$ classes, $killedClasses$ is the set of killed classes of mutant $m_i$ by $T$, and $M_e$ is the set of effective mutants.}

\black{To evaluate the impact of mutating different circuit regions and operator strategies on model performance,}
we define two metrics: \textit{KillabilityRate} (KR) and \textit{NontrivialityRate} (NR), which denote the proportion of killable or nontrivial mutants among all mutants, respectively.
\black{Under a fixed test suite\footnote{On the other hand, by using a fixed set of mutants, KR and NR can also be used to assess the quality of test suites. Since the following experiments mainly focus on evaluating the model sensitivity, we keep the test suite fixed and vary the mutants.},} 
\black{a higher KR or a lower NR indicates that a larger proportion of mutants exhibit statistically significant performance deviations from the original model, generally suggesting a stronger mutation impact and suboptimal model robustness.}
The mean accuracies of all effective mutants also provide additional insight into the sensitivity of different model regions to mutation. Combined with KR and NR, it allows a more comprehensive analysis of mutation effects.

\subsection{Test Data Preparation and Mutant Generation}

\subsubsection{Test Data of Varying Quality}
\label{sec:test_suite}

To evaluate the effectiveness of QuanForge in measuring test adequacy, \black{we construct suites with varying quality. As a reference, the original suite, denoted as \textit{Ori}, contains 100 random inputs per class. To focus on deviations introduced by mutations rather than on those inherent to the original model itself, as in \cite{deepmutation}, these samples are correctly classified by the original QNN, thereby improving the interpretability of mutation testing.
We then categorize these suites into two groups. 
Strong suites introduce more challenging inputs, including (1) \textit{LConf}, low-confidence inputs close to the decision boundary \cite{20emp,deepcrime}, (2) \textit{OOD}, out-of-distribution inputs drawn from unseen classes during training \cite{ood}, (3) \textit{Aug}, augmented inputs via affine transformations, random cropping, brightness adjustment, blurring, sharpening and additive Gaussian and salt-and-pepper noise and (4) \textit{Adv}, adversarial inputs produced by quantum FGSM algorithm\footnote{We estimate gradients using NES \cite{nes} for gradient-based attacks under finite shots. The step is configured as 50.} \cite{qaml}. Similar to Ori, we select inputs that remain correctly classified by the original model, i.e., the attack fails.
For OOD and Adv, half of the original inputs are replaced with abnormal ones. These suites contain naturally or artificially hard-to-classify patterns and are thus more likely to trigger misbehaviors of mutants.
}
\black{
In contrast, weak suites exhibit limited or biased distributions, including (1) \textit{HConf}, high-confidence inputs far from the decision boundary, (2) \textit{Skewed}, class-imbalanced inputs with ratios of 10:1 for binary tasks and 10:1:1 for ternary tasks, and (3) \textit{Small}, a half-size suite. Except for Small, all suites have the same size as Ori.}

\subsubsection{Configuration and Threshold Settings}
In the mutant generation algorithm, the gate percentage is dynamically adjusted during search.
The initial bounds for the binary search are set as $lb=0, ub=0.5$ since overly aggressive upper bounds like 1 will introduce a drastic impact and cause many trivial mutants during the initial search. $k_{max}$ and $i$ are set as 10. 
\textcolor{black}{For the mutation scope, the qubit range is set to include all qubits, and the depth range}
is set to \texttt{[0\%, 100\%]}, meaning that mutations can occur across the entire circuit by default. 
For thresholds, $\tau_{RSE}$ is 0.05 \black{following \cite{muff}}, and $\tau_{trivial}$ is 75\% \black{following \cite{deepmutation} to filter out mutants whose performance degrade significantly}. \black{The effects of threshold configurations are discussed in Section \ref{sec:param_choice}.} Finally, for the PF mutation operator, noise $x$ is sampled from a Gaussian distribution $\mathcal{N}$ with a mean of 0 and a variance of 1.

\subsection{RQ1: Effectiveness in Assessing Test Data Quality}
\label{sec:RQ1}

\begin{table}[t]
\centering
\caption{\black{Average mutation scores over all MOs obtained from the original, strong, and weak test suites. \underline{Underline} and \textbf{Bold} indicate the lowest and highest scores for a QNN among all suites, respectively.}}
\label{tab:data_quality}
\resizebox{0.8\columnwidth}{!}{%
\tiny
\begin{tabular}{@{}c|l|cccc@{\hspace{8pt}}cccc@{}}
\toprule
\multirow{2}{*}{Dataset} & \multicolumn{1}{c|}{\multirow{2}{*}{QNN}} & \multirow{2}{*}{Ori} &
\multicolumn{3}{c}{Weak Group} & \multicolumn{4}{c}{Strong Group} \\
\cmidrule(lr){4-6}\cmidrule(lr){7-10}
 & \multicolumn{1}{c|}{} &  & HConf & Skewed & Small & LConf & OOD & Aug & Adv \\ \midrule
\multirow{6}{*}{MNIST} & QCL & 0.7592 & \underline{0.5523} & 0.6958 & 0.6656 & 0.8369 & 0.8474 & \textbf{0.8921} & 0.8469 \\
 & QCNN & 0.7327 & \underline{0.4857} & 0.6292 & 0.6433 & 0.7816 & 0.7938 & \textbf{0.8419} & 0.7675 \\
 & HCQC & 0.6523 & \underline{0.3081} & 0.5101 & 0.5491 & 0.7435 & \textbf{0.7786} & 0.7714 & 0.7612 \\
 & DRNN & 0.6251 & \underline{0.4506} & 0.5905 & 0.5536 & 0.7016 & 0.7654 & 0.6953 & \textbf{0.8249} \\
 & QCL-ternary & 0.6123 & \underline{0.5226} & 0.5782 & 0.5859 & 0.6595 & \textbf{0.6835} & 0.6775 & 0.6714 \\
 & QCNN-ternary & 0.7425 & \underline{0.4658} & 0.6535 & 0.6633 & 0.7618 & \textbf{0.7739} & 0.7614 & 0.7158 \\ \midrule
\multirow{6}{*}{FashionMNIST} & QCL & 0.6445 & \underline{0.3999} & 0.5655 & 0.5657 & 0.6574 & 0.7049 & 0.6805 & \textbf{0.7185} \\
 & QCNN & 0.5879 & \underline{0.3070} & 0.5362 & 0.5848 & 0.6221 & \textbf{0.6810} & 0.6418 & 0.6551 \\
 & HCQC & 0.6469 & \underline{0.3008} & 0.6159 & 0.6211 & 0.6960 & 0.6243 & \textbf{0.7009} & 0.6613 \\
 & DRNN & 0.6321 & \underline{0.3531} & 0.5534 & 0.5839 & 0.6602 & 0.7404 & 0.6893 & \textbf{0.7789} \\
 & QCL-ternary & 0.5587 & \underline{0.3490} & 0.4926 & 0.5094 & 0.5931 & 0.6073 & 0.6009 & \textbf{0.6084} \\
 & QCNN-ternary & 0.5124 & \underline{0.3054} & 0.4541 & 0.4786 & 0.5647 & 0.5024 & 0.5648 & \textbf{0.5745} \\ \bottomrule
\end{tabular}%
}
\end{table}

Test data are essential for evaluating model performance and improving robustness, ultimately determining the generalizability and reliability of neural networks in real-world applications. High-quality test suites are typically more diverse and more likely to expose model imperfections. Since mutation operators introduce faults that alter the decision boundary, inputs near the original boundary are particularly sensitive and prone to revealing faults. Consequently, high-quality test suites are expected to achieve higher mutation scores.
\black{To validate this hypothesis, we employ the eight test suites constructed in Section \ref{sec:test_suite} to evaluate the effectiveness of QuanForge in assessing data quality.}
\black{Specifically, we generate effective mutants using different MOs, and compute the average mutation scores across all MOs for all suites, respectively.}

\black{The experimental results in Table \ref{tab:data_quality} show several observations.}
\black{\textbf{(1)} Taking the original suite as the baseline, all strong suites have consistently achieved higher mutation scores while weak suites have lower scores, which aligns with expectations. Strong suites are more sensitive to the shifts of the decision boundary brought by mutations, therefore equipped with a strong ability to kill mutants. The opposite holds for the weak suites.}
\black{\textbf{(2)} Within the weak suites, HConf suites attain the lowest scores, despite their more uniform distributions than Skewed ones and larger scales than Small ones. This is attributed to the fact that Skewed and Small suites still contain some samples near the boundary that possess the potential to kill mutants, whereas HConf is constructed by filtering for highly confident predictions, which are less sensitive to decision boundaries.}
\black{\textbf{(3)} Among the strong suites, LConf suites yield slightly lower scores than the other three suites in many cases. This is intuitive since LConf samples are naturally located far from the boundary, while the other suites are artificially strengthened by injected noise or crafted perturbations.}

Overall, results indicate a positive correlation between test suite quality and mutation score across all operators and models, demonstrating the effectiveness of QuanForge in distinguishing the quality of test data. \black{To improve the quality of test suites and expose more model misbehaviors, developers can include more inputs near the decision boundary. In practice, these samples could be collected either naturally, e.g., ambiguous samples with low-confidence predictions, or artificially, e.g., by injecting lightweight noise or adversarial perturbations. Moreover, since the strong suite that achieves the highest score can vary across different QNNs, it may be beneficial to prioritize suites with higher scores.} 
\black{Also, for model security, the mutation-based approach shows promise for detecting adversarial and backdoor samples \cite{wang2019adversarial} which tend to locate abnormally.}

\begin{center}
\begin{minipage}{0.95\linewidth}
\begin{shaded}
\noindent \textbf{Answer to RQ1:}
QuanForge is effective in evaluating the quality of the test data.
Test suites with higher quality and diversity consistently achieve higher mutation scores.
\end{shaded}
\end{minipage}
\end{center}

\subsection{RQ2: Sensitivity of Different Regions in QNNs to Mutation}

\begin{figure}[t]
    \centering
    \subfigure[\black{Effect of mutating different circuit depths on performance.}]{%
        \label{fig:depth_range_mnist}
        \includegraphics[width=0.82\textwidth]{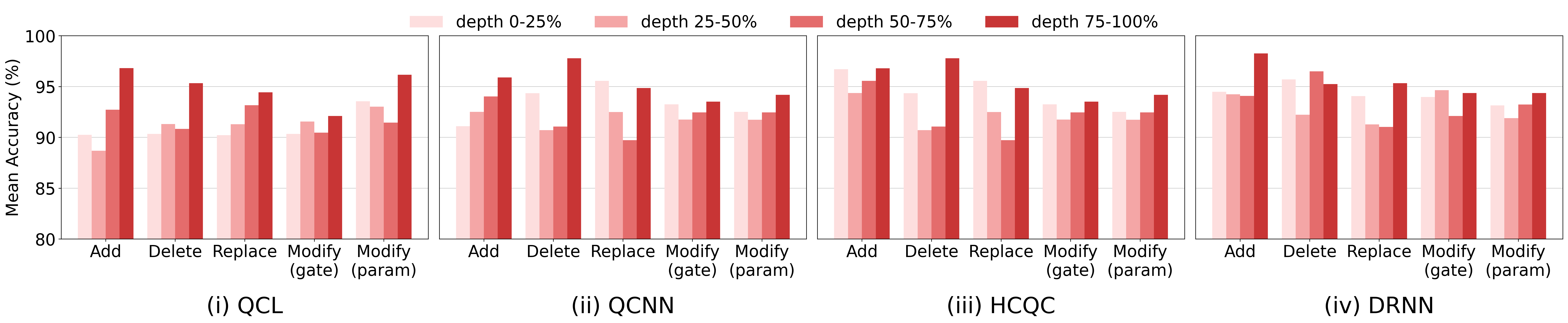}
        }
    \subfigure[\black{Effect of mutating different qubits on performance.}]{%
        \label{fig:qubit_range_mnist}
        \includegraphics[width=0.82\textwidth]{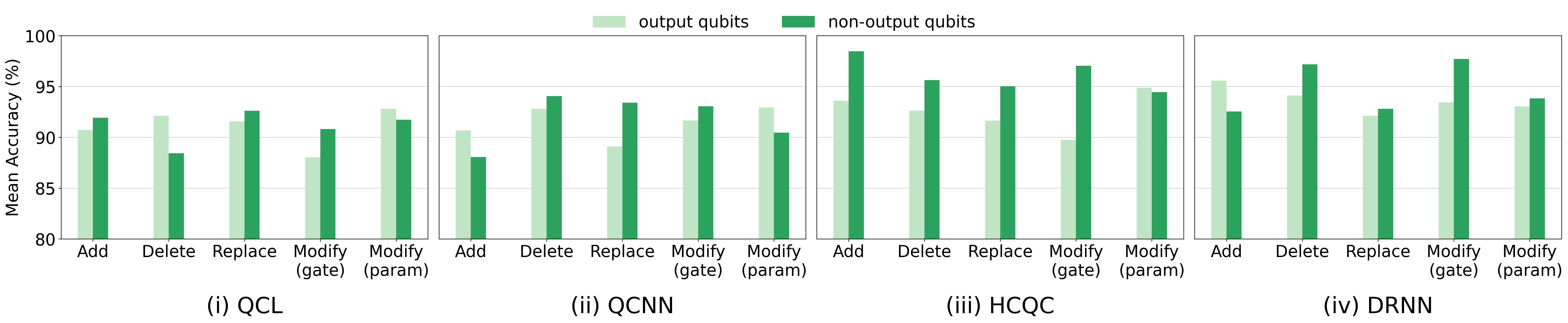}
        }
    \vspace{-4mm}
    \caption{\black{Mean accuracy of effective mutants generated over different circuit regions (MNIST).}}
    \label{fig:rq2}
    \vspace{-3mm}
\end{figure}

\begin{table}[t]
\caption{KillabilityRate and NontrivialityRate of mutants at different depths (MNIST).}
\label{tab:depth_range_mnist}
\resizebox{0.75\textwidth}{!}{
\tiny
\begin{tabular}{@{}cl|cc|cc|cc|cc@{}}
\toprule
\multirow{2}{*}{operator} & \multicolumn{1}{c|}{\multirow{2}{*}{depth range}} & \multicolumn{2}{c|}{QCL} & \multicolumn{2}{c|}{QCNN} & \multicolumn{2}{c|}{HCQC} & \multicolumn{2}{c}{DRNN} \\ \cmidrule(l){3-10} 
 & \multicolumn{1}{c|}{} & KR (\%) & NR (\%) & KR (\%) & NR (\%) & KR (\%) & NR (\%) & KR (\%)  & NR (\%)  \\ \midrule
\multirow{2}{*}{Add} & 0\%-50\% & 86.61 & 42.56 & 76.80 & 53.50 & 74.54 & 80.33 & 98.90 & 51.63 \\
 & 50\%-100\% & 85.71 & 71.55 & 79.63 & 62.38 & 65.44 & 81.84 & 93.72 & 64.29 \\ \midrule
\multirow{2}{*}{Delete} & 0\%-50\% & 90.75 & 51.50 & 92.48 & 45.68 & 86.64 & 50.36 & 98.20 & 59.88 \\
 & 50\%-100\% & 90.50 & 53.50 & 78.88 & 63.97 & 75.90 & 58.54 & 85.45 & 70.72 \\ \midrule
\multirow{2}{*}{Replace} & 0\%-50\% & 86.15 & 41.15 & 85.79 & 45.35 & 79.02 & 53.38 & 89.11 & 50.02 \\
 & 50\%-100\% & 70.48 & 68.47 & 93.57 & 63.51 & 84.62 & 56.70 & 89.06 & 66.56 \\ \midrule
\multirow{2}{*}{Modify (gate)} & 0\%-50\% & 91.26 & 39.41 & 90.57 & 45.07 & 84.46 & 56.30 & 97.82 & 48.64 \\
 & 50\%-100\% & 84.37 & 43.26 & 88.23 & 49.77 & 87.58 & 57.58 & 97.27 & 70.36 \\ \midrule
\multirow{2}{*}{Modify (param)} & 0\%-50\% & 91.96 & 46.31 & 89.62 & 47.38 & 84.57 & 60.11 & 97.75 & 69.16 \\
 & 50\%-100\% & 86.03 & 63.53 & 92.26 & 61.78 & 81.27 & 62.16 & 97.50 & 69.18 \\ \bottomrule
\end{tabular}}
\end{table}

In addition to measuring test quality, mutation testing can also probe internal model behaviors by targeting specific regions of the network~\cite{munn}. Inspired by this, here we investigate whether QuanForge can serve a similar role in analyzing quantum circuits and assessing structural robustness. Specifically, we vary the scope, depth, and qubit ranges of the mutation to systematically scan different parts of the circuit. This enables a fine-grained analysis of how mutations at various depths or on different qubits affect QNN performance. 
We report the mean accuracy of all effective mutants, KillabilityRate, and NontrivialityRate. \black{As mentioned before, after mutating different parts, we adopt \textit{the same test suite} to compute metrics, with focus on differences in model robustness. Generally, a higher KR or a lower NR indicates that more killable or trivial mutants are generated and thus the model is less robust to the target region.} 
For presentation, we categorize MOs as Add, Delete, Replace, Modify (gate), and Modify (param), where Modify (gate) denotes the average results of all gate-level modification operators (i.e., GPM, GSM, and GRT) and Modify (param) denotes the average results of all parameter-level operators.

\subsubsection{\black{RQ2.1.} Mutation at Different Circuit Depths}
\label{sec:RQ2_1}

\black{The mean accuracies on MNIST are shown in Figure \ref{fig:depth_range_mnist}.} By varying the depth range on the same QNN, we observe that mutations introduced at shallow depths (0--50\%) tend to produce greater performance degradation compared to deeper ones (75--100\%), \black{as indicated by the lower accuracies at shallow depths in general}. \black{For example, on QCL with RGA operator, shallow mutation at 0--25\% reduces accuracy to 90.25\% (a 9.75\% drop), whereas deep mutation at 75--100\% causes only a 3.18\% drop.} \black{As shown in Table \ref{tab:depth_range_mnist}, shallow-depth mutations achieve higher KRs and lower NRs, meaning that more killable or trivial mutants are generated and thus have a greater impact on the model performance.}

This observation is consistent with previous findings in DNNs~\cite{munn}, where shallow layers are crucial for early feature extraction. In QNNs, shallow-depth gates similarly extract coarse-grained features and establish early quantum correlations. As the circuit deepens, additional transformations build on this foundation, propagating information through more complex interactions. Thus, early mutations are amplified through the circuit, whereas deep mutations exert more localized effects.

\subsubsection{\black{RQ2.2.} Mutation at Different Qubits}
\label{sec:RQ2_2}

\begin{table}[t]
\caption{KillabilityRate and NontrivialityRate of mutants targeting at output and non-output qubits (MNIST).}
\label{tab:qubit_range_mnist}
\resizebox{0.75\textwidth}{!}{
\tiny
\begin{tabular}{@{}cl|cc|cc|cc|cc@{}}
\toprule
\multirow{2}{*}{operator} & \multicolumn{1}{c|}{\multirow{2}{*}{qubit range}} & \multicolumn{2}{c|}{QCL} & \multicolumn{2}{c|}{QCNN} & \multicolumn{2}{c|}{HCQC} & \multicolumn{2}{c}{DRNN} \\ \cmidrule(l){3-10} 
 & \multicolumn{1}{c|}{} & KR (\%) & NR (\%) & KR (\%) & NR (\%) & KR (\%) & NR (\%) & KR (\%) & NR (\%)  \\ \midrule
\multirow{2}{*}{Add} & output & 89.27 & 40.98 & 76.86 & 50.98 & 69.61 & 66.67 & 94.29 & 52.65 \\
 & non-output & 85.79 & 44.21 & 57.00 & 59.40 & 58.18 & 85.16 & 91.67 & 65.83 \\ \midrule
\multirow{2}{*}{Delete} & output & 92.96 & 41.85 & 96.36 & 42.95 & 80.88 & 45.96 & 94.34 & 43.68 \\
 & non-output & 96.67 & 34.33 & 82.97 & 45.13 & 84.81 & 62.22 & 86.00 & 100 \\ \midrule
\multirow{2}{*}{Replace} & output & 82.75 & 40.75 & 92.18 & 49.38 & 57.33 & 64.67 & 97.55 & 40.41 \\
 & non-output & 78.95 & 54.79 & 87.20 & 48.40 & 53.33 & 82.08 & 90.56 & 91.11 \\ \midrule
\multirow{2}{*}{Modify (gate)} & output & 86.50 & 37.00 & 91.79 & 48.93 & 79.63 & 41.09 & 96.91 & 41.27 \\
 & non-output & 93.75 & 32.75 & 92.55 & 35.74 & 79.11 & 72.05 & 95.78 & 100 \\ \midrule
\multirow{2}{*}{Modify (param)} & output & 89.41 & 52.56 & 92.22 & 54.07 & 68.57 & 73.93 & 98.89 & 58.52 \\
 & non-output & 93.64 & 56.00 & 93.22 & 50.32 & 88.25 & 76.00 & 100 & 100 \\ \bottomrule
\end{tabular}}
\end{table}

In current QNN designs, model outputs are typically defined by the expectation values measured on specific qubits.
It is intuitive to consider that mutating these output qubits might have a more direct and significant impact on model behavior, \black{causing mutants more likely to deviate from the original model.} To investigate this, we \black{consider two settings for qubit range, i.e., qubits used as QNN outputs (e.g., qubits 0 and 1 for a binary task) and an equally sized random subset selected from the remaining qubits (e.g., qubits 2 and 5).} The depth range is fixed to \texttt{[0\%, 100\%]} here. Results of mean accuracies and two rates are shown in Figure \ref{fig:qubit_range_mnist} and Table \ref{tab:qubit_range_mnist}, respectively.

\black{The experimental results indicate that the sensitivity to output-qubit mutations depends on specific circuit structures. For QCL and QCNN, the effect of such mutations varies across operators without a consistent pattern. In some cases, mutations on non-output qubits cause even greater degradation (e.g., QCL with RGD). In contrast, HCQC and DRNN are more sensitive to output-qubit mutations, resulting in lower mutant accuracy for most operators. This trend is also reflected in NRs. For instance, on DRNN with Delete, output-qubit mutations yield a NR of 43.68\% compared to that of 100\% for non-output mutations, indicating that output-qubit changes disturb the decision boundary more easily and generate more trivial mutants. 
The difference arises mainly from the circuit scale and encoding methods. HCQC adopts a small-scale circuit with a single output qubit (see Table \ref{tab:dataset}), where limited entanglement concentrates information on the output qubit, amplifying mutation effects. The sensitivity of DRNN comes from angle encoding. It maps each input feature to qubit rotations with a constant circuit depth \cite{qml2}, so output-qubit mutations directly distort encoded states. In contrast, amplitude encoding (used in QCL and QCNN) distributes features globally across $2^n$ basis states through an exponentially deep circuit \cite{hcqc}, making few-qubit mutations less impactful. Their deeper, highly entangled circuits also spread local perturbations across qubits, narrowing the gap between output- and non-output-qubit mutations.}

\begin{figure}[t]
    \centering
    \includegraphics[width=0.8\linewidth]{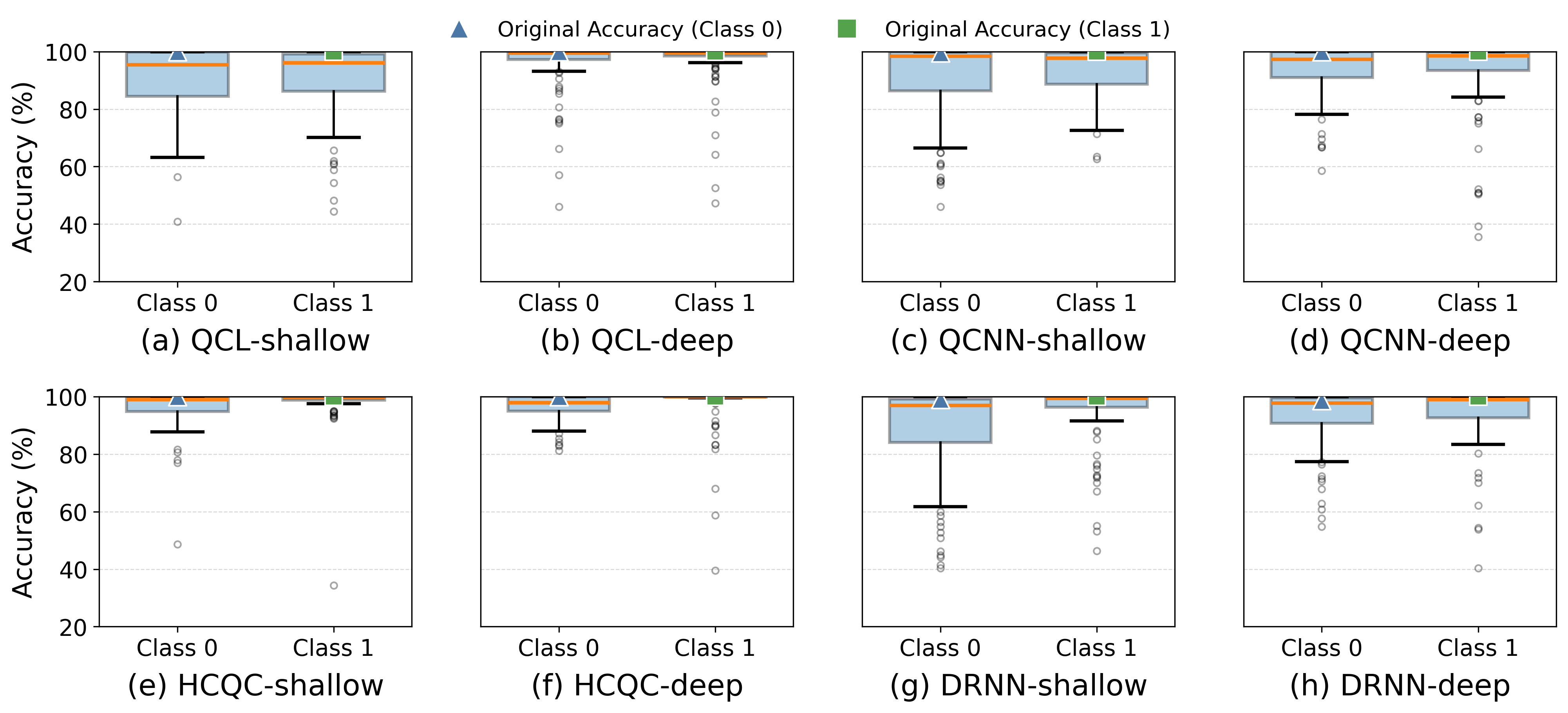}
    \vspace{-3mm}
    \caption{\black{Accuracy distribution of mutants generated by MNIST and RGA operator at different depths.}}
    \label{fig:mut_acc_dis}
\end{figure}

\black{After identifying the sensitive regions, we further examine what kinds of errors are induced by mutations by collecting per-class accuracies of all mutants. Figure \ref{fig:mut_acc_dis} indicates that shallow-depth mutants exhibit more severe accuracy degradation on class 0, suggesting an output bias towards class 1. The bias still exists, but is less pronounced for deep-depth mutants.} 

Generally, these findings suggest that QuanForge can be used not only for testing data quality but also for analyzing model internals. Vulnerable regions identified by specified-region mutation could inform \black{future model development and enhancement.}
\black{
Other techniques in classical models, including fault localization~\cite{mt4faultlocate}, weight ranking~\cite{genmunn}, and performance recovery~\cite{care}, could also be extended to quantum circuits in future research.
From a quantum-specific perspective, the ansatz architecture design (e.g., more trainable gates) within these regions can be refined to investigate their correlation with model expressivity \cite{du2022efficient} and generalization \cite{yang2025stability}.}

\begin{center}
\begin{minipage}{0.95\linewidth}
\begin{shaded}
\noindent \textbf{Answer to RQ2:}
Mutations at shallow depths have a greater impact on QNN performance, as early-stage gates extract foundational features whose perturbations propagate and amplify through the circuit. The sensitivity of QNNs to qubit-level mutations is affected by the specific circuit scale, encoding schema and entanglement complexity.
\end{shaded}
\end{minipage}
\end{center}

\subsection{RQ3: Impact of Mutating Different Quantum Gates on Model Performance}
\label{sec:RQ3}

\begin{table}[t]
\caption{Evaluation results of mutants targeted at different gate types. Acc, KR, and NR denote mean accuracy, KillabilityRate, and NontrivialityRate, respectively. 'N/A' denotes the cases where this type of gate does not exist, or the MO does not apply to the gates in the given QNN. \underline{Underline} indicates the highest sensitivity across different gate types for a specific QNN and operator.}
\label{tab:gate_type}
\resizebox{0.85\textwidth}{!}{
\tiny
\begin{tabular}{@{}cl|ccc|ccc|ccc|ccc@{}}
\toprule
\multirow{2}{*}{operator} & \multicolumn{1}{c|}{\multirow{2}{*}{gate type}} & \multicolumn{3}{c|}{QCL} & \multicolumn{3}{c|}{QCNN} & \multicolumn{3}{c|}{HCQC} & \multicolumn{3}{c}{DRNN} \\ \cmidrule(l){3-14} 
 & \multicolumn{1}{c|}{} & Acc & KR & NR & Acc & KR & NR & Acc & KR & NR & Acc & KR & NR \\ \midrule
\multirow{6}{*}{Add} & Controlled & {\underline{88.17}} & 89.63 & 46.67 & 92.73 & 91.11 & 52.59 & 94.67 & {\underline{95.76}} & 67.88 & 94.92 & 95.00 & 59.17 \\
 & Hadamard & 88.58 & {\underline{94.86}} & 45.14 & {\underline{90.8}} & 89.75 & 63.16 & {\underline{90.09}} & 77.29 & 54.17 & {\underline{94.03}} & {\underline{98.00}} & {\underline{42.25}} \\
 & Pauli & 91.68 & 88.89 & {\underline{36.19}} & 92.44 & 82.41 & 61.35 & 95.63 & 58.52 & 82.96 & 94.63 & 96.82 & 47.05 \\
 & Phase & 99.95 & 26.00 & 100 & 94.01 & 94.14 & 68.97 & 95.87 & 84.14 & 81.38 & 99.81 & 85.00 & 100 \\
 & Rotation & 97.72 & 93.68 & 61.58 & 93.78 & 90.00 & 61.03 & 96.19 & 85.00 & 75.29 & 99.68 & 76.00 & 100 \\
 & Swap & 89.77 & 92.08 & 37.92 & 92.20 & {\underline {95.81}} & {\underline{41.61}} & 94.88 & 89.42 & {\underline{51.15}} & 91.56 & 95.88 & 47.35 \\ \midrule
\multirow{3}{*}{Delete} & Controlled & {\underline {91.14}} & {\underline{100}} & {\underline{35.42}} & 91.45 & {\underline{97.39}} & {\underline{46.21}} & 97.00 & {\underline{94.19}} & {\underline{60.00}} & 94.74 & 95.56 & 70.31 \\
 & Pauli & N/A & N/A & N/A & {\underline{87.87}} & 50.95 & 70.95 & N/A & N/A & N/A & N/A & N/A & N/A \\
 & Rotation & 94.45 & 94.41 & 40.88 & 94.63 & 83.57 & 48.81 & {\underline{94.13}} & 85.34 & 64.65 & {\underline{93.21}} & {\underline{95.71}} & {\underline{64.76}} \\ \midrule
\multirow{3}{*}{Replace} & Controlled & {\underline {88.57}} & \underline{87.33} & 55.33 & {\underline{92.40}} & {\underline{84.87}} & {\underline{51.15}} & {\underline{93.53}} & 75.28 & {\underline{58.82}} & 93.44 & {\underline{96.67}} & 57.04 \\
 & Pauli & N/A & N/A & N/A & 97.71 & 65.00 & 87.31 & N/A & N/A & N/A & N/A & N/A & N/A \\
 & Rotation & 92.45 & 74.17 & \underline{49.79} & 92.49 & 84.70 & 51.18 & 93.67 & {\underline{78.82}} & 59.17 & {\underline{92.97}} & 96.51 & {\underline{51.86}} \\ \midrule
\multirow{3}{*}{Modify (gate)} & Controlled & {\underline{89.09}} & {\underline{99.43}} & {\underline{24.00}} & {\underline{92.47}} & {\underline {91.45}} & {\underline{39.64}} & {\underline{94.92}} & {\underline{93.32}} & {\underline{59.60}} & 94.16 & 94.61 & 59.23 \\
 & Pauli & N/A & N/A & N/A & 93.08 & 86.15 & 46.03 & N/A & N/A & N/A & N/A & N/A & N/A \\
 & Rotation & 91.59 & 82.17 & 44.35 & 93.47 & 84.03 & 48.38 & 95.31 & 77.36 & 62.60 & {\underline{93.64}} & {\underline{96.17}} & {\underline{50.00}} \\ \midrule
\multirow{3}{*}{Modify (param)} & Controlled & N/A & N/A & N/A & 97.85 & 64.69 & 89.07 & N/A & N/A & N/A & 97.41 & 96.67 & 95.56 \\
 & Pauli & N/A & N/A & N/A & N/A & N/A & N/A & N/A & N/A & N/A & N/A & N/A & N/A \\
 & Rotation & 93.20 & 95.20 & 57.20 & {\underline{94.38}} & {\underline{92.98}} & {\underline{64.46}} & 93.80 & 82.80 & 61.00 & {\underline{94.78}} & {\underline{100}} & {\underline{83.89}} \\ \bottomrule
\end{tabular}}
\end{table}

Quantum gates differ in their functional roles. Some gates, such as Hadamard and controlled gates, have strong, rigid effects that significantly alter entanglement or transform basis states. Others, like phase gates, apply more subtle modifications. To analyze how gate types influence model behavior under mutation, we configure the parameter gate type for each mutation operator and perform controlled mutations. Note that the Add operator can insert gates of any specified type, while other operators can only mutate gates already present in the circuit. \black{For Delete, Replace, and Modify operators, we reduce the initial $ub$ of the search algorithm to ensure that the number of target gates remains comparable in early-stage mutants under different gate types. This is because QNNs usually contain an imbalanced distribution of different gate types, with rotation gates typically occupying a larger proportion. Hence, mutants could disproportionally involve rotation gates under the same configuration compared to other types, leading to unfair comparison.} \black{Also, similar to RQ2, we use the same test suite to compute metrics for the same QNN and MO.} 

The results in Table \ref{tab:gate_type} show that different gate types have markedly different impacts on model performance, depending on both the gate category and the operator applied.
For the Add operator, adding Hadamard and SWAP gates tends to cause the most significant performance degradation, in general, \black{indicated by lower mean accuracies, higher KRs, and lower NRs}. To explain this, the Hadamard gate transforms a basis state into a superposition state, introducing abrupt entanglement. The SWAP gate, which exchanges the states of two qubits, also ranks high in terms of impact. In contrast, phase and rotation gates have relatively mild effects. \black{The limited impact of rotation gates is partially attributed to the inherent robustness of quantum circuits to noise \cite{noiseforrobust},} while phase gates modify the relative phase of a state but do not affect its amplitude, which is less tied to the measurement results.

For operators that delete or modify existing gates, amplitude-encoding QNNs tend to be more sensitive to controlled gates. This behavior is primarily due to the entangling nature of these gates. Altering or removing them disrupts multi-qubit correlations, which are essential to the QNN's ability to encode and process information. Since these correlations are distributed non-locally, such mutations often induce significant changes in model behavior. \black{A special case is QCNN with Modify (gate) operators, where QCNN is more sensitive to rotation gates. This is because rotation gates in QCNN are implemented as U3 gates, whose all 3 parameters are subject to mutation, thus causing a stronger impact on the model.} \black{Angle-encoding DRNN is also sensitive to rotation gates. In DRNN, these gates serve as both feature encoders and extractors, playing a more dominant role in the circuit. Moreover, DRNN adopts the CRZ gate as a controlled operation, which, according to our categorization, also belongs to the rotation type.}


\black{The observed sensitivity to different quantum gates also provides practical implications for circuit design. Developers could adjust the positions and counts of specific gate types and assess their relationships with model properties, including expressivity, trainability, and robustness.}

\begin{center}
\begin{minipage}{0.95\linewidth}
\begin{shaded}
\noindent \textbf{Answer to RQ3:}
For the Add operator, the Hadamard and SWAP gates cause the most significant performance degradation due to their strong, non-local transformations. In contrast, phase and rotation gates have milder effects. For the Delete and Modify operators, controlled gates have a high impact because they disrupt entanglement and non-local correlations.
\end{shaded}
\end{minipage}
\end{center}

\subsection{Discussion}

\subsubsection{Performance of different operators in generating effective mutants}

\begin{table}[t]
\caption{Comparison between MOs in the capability of generating killable and nontrivial mutants (MNIST). On each QNN, \textbf{bold} indicates the highest KillabilityRate or NontrivialityRate across all MOs.}
\label{tab:mo}
\resizebox{0.7\textwidth}{!}{
\tiny
\begin{tabular}{@{}c|c|cc|cc|cc|cc@{}}
\toprule
\multirow{2}{*}{type} & \multirow{2}{*}{name} & \multicolumn{2}{c|}{QCL} & \multicolumn{2}{c|}{QCNN} & \multicolumn{2}{c|}{HCQC} & \multicolumn{2}{c}{DRNN} \\ \cmidrule(l){3-10} 
 &  & KR (\%) & NR (\%) & KR (\%) & NR (\%) & KR (\%) & NR (\%) & KR (\%) & NR (\%) \\ \midrule
Add & RGA & 83.49 & 47.44 & 69.39 & 56.12 & 68.95 & \textbf{68.68} & 90.95 & 54.05 \\
Delete & RGD & 94.32 & 41.35 & \textbf{94.38} & 45.83 & 84.20 & 53.40 & 96.18 & 52.56 \\
Replace & GR & 87.11 & 41.84 & 85.85 & 51.22 & 76.04 & 55.81 & 88.79 & 53.33 \\ \midrule
\multirow{3}{*}{Modify (gate)} & GPM & 96.94 & 37.50 & 87.41 & 46.29 & 80.82 & 53.15 & 91.89 & 45.68 \\
 & GSM & 91.84 & 48.42 & 87.41 & 46.30 & 87.14 & 55.71 & 97.84 & 41.35 \\
 & GRT & 91.05 & 44.47 & 92.29 & 36.25 & 86.44 & 53.56 & 92.55 & 42.34 \\ \midrule
\multirow{3}{*}{Modify (param)} & PF & \textbf{98.75} & \textbf{53.75} & 93.92 & \textbf{61.79} & 81.75 & 61.75 & 98.82 & \textbf{71.18} \\
 & PSF & 91.74 & 41.96 & 83.58 & 50.75 & 80.00 & 64.47 & 98.92 & 46.22 \\
 & PS & 84.20 & 46.80 & 91.14 & 37.14 & \textbf{89.47} & 51.84 & \textbf{99.05} & 42.86 \\ \bottomrule
\end{tabular}}
\end{table}

\black{Mutation operators affect QNNs through heterogeneous faults, introducing noisy transformations, loss of entanglement, and changes in transformation flow. As a result, their influence on model performance differs. Aggressive operators are more likely to generate trivial mutants with lower NRs, while moderate ones tend to produce more unkillable mutants with lower KRs. Here, we investigate the generation capabilities of different operators under the same settings as in RQ1. The KRs and NRs of different operators are reported in Table \ref{tab:mo}.}

\black{We observe that the Add operator is relatively mild in mutation strength. It generally exhibits a lower KR and a higher NR. This finding is consistent with a previous study \cite{mutationoperator}.}
\black{Currently, QuanForge adopts a single MO per mutant. Based on the performance of MOs, future work can explore combinations of different MOs to produce mutants with greater complexity and diversity.} \black{Moreover, a prioritization strategy for MOs could be developed to improve the generation efficiency of effective mutants, e.g., reducing the weight of MOs that tend to produce trivial mutants.}

\subsubsection{Sensitivity to parameter choice}
\label{sec:param_choice}

\begin{table}[t]
\centering
\caption{\black{Effective mutants, mutation scores, and region sensitivity under different $\tau_{RSE}$ and $\tau_{trivial}$.}}
\label{tab:param_choice}
\resizebox{0.95\columnwidth}{!}{%
\tiny
\begin{tabular}{@{}c|cc|ccccc|cccccc@{}}
\toprule
\multirow{2}{*}{} & \multicolumn{2}{c|}{\multirow{2}{*}{\begin{tabular}[c]{@{}c@{}}Parameter\\ configurations\end{tabular}}} & \multirow{2}{*}{\begin{tabular}[c]{@{}c@{}}\# Effective \\ mutants \\ number \end{tabular}} & \multicolumn{4}{c|}{Mutation score} & \multicolumn{2}{c}{KR (\%)} & \multicolumn{2}{c}{NR (\%)} & \multicolumn{2}{c}{Acc (\%)} \\ \cmidrule(l){5-14} 
 & \multicolumn{2}{c|}{} &  & Ori & HConf & LConf & Aug & \begin{tabular}[c]{@{}c@{}}depth\\ 0-25\%\end{tabular} & \begin{tabular}[c]{@{}c@{}}depth\\ 75-100\%\end{tabular} & \begin{tabular}[c]{@{}c@{}}depth\\ 0-25\%\end{tabular} & \begin{tabular}[c]{@{}c@{}}depth\\ 75-100\%\end{tabular} & \begin{tabular}[c]{@{}c@{}}depth\\ 0-25\%\end{tabular} & \begin{tabular}[c]{@{}c@{}}depth\\ 75-100\%\end{tabular} \\ \midrule
\multirow{4}{*}{\begin{tabular}[c]{@{}c@{}}MNIST\\ and\\ QCL\end{tabular}} & \multirow{4}{*}{$\tau_{RSE}$} & 1\% & 0 & - & - & - & - & - & - & - & - & - & - \\
 &  & 3\% & 97 & 0.5795 & 0.3636 & 0.7841 & 0.8295 & 50.80 & 50.83 & 80.81 & 90.83 & 90.97 & 94.26 \\
 &  & 5\% & 75 & 0.7312 & 0.3855 & 0.8067 & 0.8667 & 80.69 & 79.56 & 38.60 & 85.21 & 89.99 & 95.87 \\
 &  & 10\% & 23 & 0.6739 & 0.5000 & 0.7174 & 0.7826 & 89.33 & 88.41 & 43.33 & 56.89 & 89.98 & 93.15 \\ \midrule
\multirow{4}{*}{\begin{tabular}[c]{@{}c@{}}FashionMNIST\\ and\\ QCL\end{tabular}} & \multirow{4}{*}{$\tau_{trivial}$} & 60\% & 111 & 0.7388 & 0.5991 & 0.7982 & 0.7928 & 64.06 & 63.10 & 62.19 & 87.24 & 78.61 & 90.91 \\
 &  & 75\% & 81 & 0.5274 & 0.3642 & 0.5432 & 0.6111 & 70.93 & 70.63 & 46.56 & 75.59 & 88.49 & 92.40 \\
 &  & 90\% & 42 & 0.6667 & 0.2500 & 0.6547 & 0.7381 & 72.50 & 68.00 & 40.94 & 59.14 & 95.12 & 97.13 \\
 &  & 95\% & 23 & 0.6074 & 0.2174 & 0.7174 & 0.6521 & 74.19 & 61.84 & 38.06 & 61.32 & 95.95 & 97.23 \\ \bottomrule
\end{tabular}%
}
\end{table}

The choices of $\tau_{RSE}$ and $\tau_{trivial}$ directly determine how strictly QuanForge enforces mutant stability and accuracy. To conduct a sensitivity analysis by varying the two thresholds across multiple configurations, respectively, we repeat the experiments in RQ1 and RQ2.1 and report the corresponding metrics.

\black{As in Table \ref{tab:param_choice}, under a smaller $\tau_{RSE}$, more effective mutants are generated since it imposes a stricter stability threshold, which requires accumulating more mutants for a smaller RSE. Note that, for some extreme settings such as 1\%, sufficiently stable mutants may even become unavailable under the given budget. 
On the other hand, a larger $\tau_{trivial}$ imposes a higher accuracy constraint, i.e., mutants should remain closer to the original model's performance, thereby reducing the number of effective mutants. Despite quantitative changes in the metrics, the main findings remain consistent with the original experiments across configurations. Strong suites still achieve higher scores than the weak ones. Also, shallow-depth mutations manifest higher KRs, lower NRs, and lower accuracies than deep-depth ones.} 
\black{To ensure a sufficient yield of effective mutants, we recommend avoiding overly strict settings like very small $\tau_{RSE}$ or very large $\tau_{trivial}$.}

\subsubsection{Evaluation on noisy models}

\begin{table}[t]
\centering
\caption{\black{Mutation scores of different test suites for MNIST and noisy HCQC\_TTN.}}
\label{tab:noisy}
\resizebox{0.65\columnwidth}{!}{%
\tiny
\begin{tabular}{@{}c|cccccccc@{}}
\toprule
\multirow{2}{*}{operator} & \multirow{2}{*}{Ori} & \multicolumn{3}{c}{Weak group} & \multicolumn{4}{c}{Strong group} \\
\cmidrule(lr){3-5}\cmidrule(lr){6-9}
 &  & HConf & Skewed & Small & LConf & OOD & Aug & Adv \\ \midrule
Add & 0.7701 & {\underline{0.3834}} & 0.5822 & 0.7123 & 0.9112 & 0.7923 & 0.7922 & \textbf{0.9403} \\
Delete & 0.6046 & {\underline{0.1792}} & 0.5633 & 0.5384 & 0.8823 & 0.7564 & 0.8412 & \textbf{0.8974} \\
Replace & 0.5534 & {\underline{0.1292}} & 0.5323 & 0.5104 & 0.6681 & 0.8428 & 0.8292 & \textbf{0.9514} \\
Modify (gate) & 0.7153 & {\underline{0.2500}} & 0.6633 & 0.5400 & 0.8533 & 0.7833 & 0.8567 & \textbf{0.9486} \\
Modify (param) & 0.5371 & {\underline{0.1736}} & 0.5914 & 0.5593 & 0.7272 & 0.7650 & 0.7657 & \textbf{0.8550} \\ \bottomrule
\end{tabular}%
}
\end{table}

Real quantum systems suffer from unwanted interactions with the outside world, appearing as noise that can distort quantum evolution and degrade execution fidelity \cite{nielsen2010quantum}.
The existence of quantum noise inevitably introduces perturbations into QNN outputs, thereby impairing the accuracy of mutation testing. To assess the robustness and effectiveness of QuanForge under noisy conditions, \black{we evaluate it against several representative noise channels on the \texttt{default.mixed} simulator of PennyLane, including depolarizing, bit flip, phase flip, phase damping, amplitude damping, crosstalk, drift, and thermal relaxation, with reference to APIs~\cite{pennylane_noise} or noise models~\cite{crosstalk_drift}.
They are inserted at random positions with a probability of 1\% to approximate real hardware conditions, and detailed parameters are selected referring to Qiskit documents \cite{noise_param1,noise_param2}.}
To alleviate computational overhead, we train a smaller-scale HCQC using the ansatz U\_TTN.
All other algorithmic and mutation operator settings are consistent with RQ1. \black{The results in Table \ref{tab:noisy} show that, compared to the original suites, strong suites have still achieved higher scores than weak ones under all operators. Under all operators, HConf and Adv suites yield the lowest and highest scores, respectively. This illustrates the effectiveness of QuanForge in distinguishing the quality of test suites under noisy quantum conditions.} 

\subsection{Threats to Validity}

\black{\textit{External Validity.} Experiments are conducted on a limited set of datasets and QNN architectures. Additional architectures and larger-scale circuits could be explored to improve generalizability and scalability. Additionally, the simulated noise may differ from noise on real quantum hardware, which is device-specific and unpredictable.} The parameter selections for different channels constitute another threat, which is alleviated by referring to examples used in open-source implementations.

\textit{Internal Validity.} We employ GLM and Cohen's $d$ to judge the mutation killing, which is commonly used in mutation testing for neural networks \cite{deepcrime,20emp}. Adversarial examples are generated using NES-based black-box gradient estimation. Alternative gradient estimators may lead to different perturbations.
Redundant or equivalent mutants are not explicitly excluded. To mitigate this risk, we generate mutants by applying the same operator to different gate positions, thereby increasing structural diversity.
Randomness in gate selection, input sampling, noise injection, and model training may introduce variability, which we mitigate by running repeated experiments and using fixed random seeds.

\section{Conclusion}
\label{sec:conclusion}

In this paper, we proposed QuanForge, a \black{post-training} mutation testing technique tailored for QNNs, to support the evaluation of both test data quality and model structural sensitivity. By introducing multi-granularity mutation operators, diverse faults are injected into parameterized quantum circuits, simulating both logical and physically realistic errors. \black{Combined with statistical mutation killing, QuanForge utilizes stability checking and effectiveness analysis to ensure the mutant quality. It addresses the challenges posed by randomness in operator application and quantum measurement.} Extensive experiments demonstrated the effectiveness of QuanForge in assessing the test data quality and revealing fragile regions in QNNs. We further analyzed \black{the sensitivity to various gate types and} the generation capability of different operators, which guide future mutation design. \black{To approach more realistic quantum execution, the effectiveness of QuanForge was finally evaluated against multiple simulated noise.}

\section*{Data Availability}
Our framework is publicly available at \url{https://github.com/MinqiShao/QuanForge}.

\section*{Acknowledgments}
This work was supported by JST SPRING Grant No.\ JPMJSP2136, JST BOOST Grant No.\ JPMJBS2406, and JSPS KAKENHI Grants No.\, JP26K02892 and No.\ JP24K14908.

\clearpage
\bibliographystyle{ACM-Reference-Format}
\bibliography{reference}

\end{document}